\documentstyle[aps,prl,preprint]{revtex}
\begin{document}
\widetext
\title{
Continuum quantum ferromagnets at finite temperature\\
and the quantum Hall effect}
\author{N. Read and Subir Sachdev}
\address{Department of Physics, P.O. Box 208120, Yale University,
New Haven, CT 06520-8120\\
and\\
Department of Applied Physics, P.O. Box 208284, Yale University,
New Haven, CT 06520-8284}
\date{July 10, 1995}
\maketitle

\widetext
\begin{abstract}
We study finite temperature ($T$) properties of the continuum quantum field
theory of
systems with a ferromagnetic ground state.
A scaling theory of the $T=0$ system is discussed carefully, and its
consequences
for crossovers between
different finite $T$ regimes in dimensions 1, 2, and 3 are described. The
results
are compared with
recent NMR measurements of the magnetization
of a quantum Hall system with filling factor $\nu=1$; we predict that the
relaxation rate $1/T_1$ of this system may have a finite $T$ ``ferromagnetic
coherence peak''.
\end{abstract}

\narrowtext

The recent availability of nuclear magnetic resonance (NMR)
measurements~\cite{sean}
of quantum Hall systems has opened a new window onto the magnetic
properties of a
strongly correlated two-dimensional electronic system. Initially, the
filling factor ($\nu$) dependence of the zero temperature ($T$)
magnetization in the
vicinity of $\nu=1$ attracted attention because it indicated that the low
energy
charged excitations of the system were spin textures (skyrmions)
\cite{kane,sondhi,fertig}.
In this paper, we examine instead the $T$ dependence of the magnetic
properties exactly at
$\nu=1$~\cite{kasner}. We use a continuum
quantum field theory of a ferromagnet as a model and describe its finite $T$
properties. From a field-theoretical perspective, some features of this
quantum field theory are rather unusual and
lead to a noteworthy universality in the crossover
functions. We will consider all values of the spatial dimension $d>0$, although
the regime of validity of the continuum limit becomes larger as $d$ is
lowered and
it is most useful for $d \leq 2$. Our theory can also be applied to
other low-dimensional ferromagnets (like the ferromagnetic layer of
${}^{3} He$ on Grafoil~\cite{osheroff}) but we will limit our discussion
here to the
quantum Hall system. Some limitations of the model as applied to the
quantum Hall effect
will also be discussed.

The required quantum field theory is obtained from the naive continuum
limit of the
coherent-state path integral of an insulating, lattice ferromagnet:
\begin{displaymath}
Z = \int \! {\cal D} \vec{n} \, \delta ( \vec{n}^2 - 1) \exp\left(\! - \!
\int \! d^d x \int_0^{1/T} \!\! d\tau ({\cal L}_0 [\vec{n}]
+ {\cal L}_1 [\vec{n}] )\right)
\end{displaymath}
\begin{equation}
{\cal L}_0 [\vec{n}] =
i M_0 \vec{A}(\vec{n}) \cdot \partial_{\tau} \vec{n}
+ (\rho_s /2) ( \nabla_x \vec{n} )^2 - M_0 \vec{H} \cdot \vec{n}.
\label{cqfm}
\end{equation}
Here $\vec{n} (x,\tau)$ is the 3-component unit vector field identifying the
local orientation of the ferromagnetic order (it is periodic in the
Matsubara time
$\tau$), and $M_0\geq 0$ is the magnetization per unit volume in the
ferromagnetic ground
state. The first term in ${\cal L}_0$ is the kinematical Berry
phase~\cite{haldane}
which accounts for the commutation relations between the components of the
order parameter; $\vec{A}$ is the vector potential of a unit Dirac monopole at
the origin of spin space with $\epsilon_{ijk} \partial A_{k}/ \partial n_j=
n_i$,
$\rho_s$ is the ground state spin stiffness, and $H$ is the magnetic field;
${\cal L}_1$
contains local higher-gradient terms, with no time derivatives, that will
be discussed below. (The Hopf term, which does contain a time
derivative, will be discussed separately later.)
We are using units in which $k_B = \hbar = 1$ and have
absorbed a factor of $g\mu_B$ into $H$ ($\mu_B$ is the Bohr magneton).
For a ferromagnet on a hypercubic lattice with spacing $a$, spin per site $S$,
and nearest-neighbor exchange $J$, $M_0 = S a^{-d}$ and $\rho_s = J S^2
a^{2-d}$.
In the quantum Hall effect at $\nu=1$, $M_0=1/(4\pi \ell_B^2)$, and
(neglecting layer finite-thickness corrections which are
expected to reduce $\rho_s$
\cite{sondhi}) $\rho_s=e^2/(16\sqrt{2\pi}\varepsilon\ell_B)$
\cite{kallin,sondhi},
where $\ell_B$ is the magnetic length.
In the experiment of Ref.~\cite{sean}, $H \approx 2 \mbox{K}$ (note that
$g\approx 0.5$
in {\it GaAs}), while we estimate that $\rho_s \approx 3 \mbox{K}$.

Since ${\cal L}_1$ contains no time derivatives,
the Hilbert space is fully determined (through canonical arguments) by $M_0$,
and the remainder of the action describes the Hamiltonian acting in this space.
For $M_0=0$, there are no degrees of freedom in the system (the Hilbert space
is one dimensional), so the Hamiltonian is immaterial.
For $M_0\neq0$, the explicit quantization of the continuum quantum ferromagnet
(CQFM) defined by Eqn~(\ref{cqfm}) is difficult, but it
is not hard to establish the quantization condition
that $2M_0 L^d$ must be integral; $M_0 L^d$ is the total spin of the
fully polarized state ($L^d$ is the volume of the system). Since all
states must have half-integral spin, we can associate a length $\xi_0$ with
$M_0$, $2M_0=\xi_0^{-d}$. We expect that the degrees of freedom of the
CQFM correspond roughly to independent spins $1/2$ per volume $\xi_0^d$, and
the scale $\xi_0$ shows up naturally in the quantum theory \cite{hald86}; for
example, it is likely that the commutation relations for the spin density
operators are smeared over the scale $\xi_0$ (a similar effect occurs in the
quantum Hall system as a result of restriction to the lowest Landau level),
and that the correlation length is never less than $\xi_0$.

In determining the applicability of the CQFM to a real system, we must consider
a renormalization group (RG) analysis.
There is a fixed point at $M_0=0$, and the terms in ${\cal L}_0$ are
the most relevant perturbations.
This can be seen by simple power counting: under a rescaling
$x\rightarrow e^{-\ell}x$, $\tau\rightarrow e^{-z\ell}\tau$ (with $z=2$ so
that the long wavelength spin wave dispersion is invariant), we find that
$M_0$ has dimension $d$ (corresponding to its $-d$ powers of $\xi_0$),
$\rho_s$ has dimension $d+z-2=d$, and $T$ and $H$ have dimension $z=2$. Terms
in ${\cal L}_1$ have $k \geq 4$ gradients, and their coefficients have
dimension $d+z-k$; for $d <2$, all such terms are irrelevant.
To go beyond power-counting requires a diagrammatic RG which will be
described elsewhere; the results include an RG reinterpretation of
earlier spin wave calculations in
$d=3$ \cite{dyson} and $d=2$ \cite{kosevich}.
At $T=0$, the power-counting flows for the couplings in ${\cal L}_0$ are exact,
but these couplings do generate a term in ${\cal L}_1$,
$\lambda (  \partial_a n_i \partial_a n_i \partial_b n_j \partial_b n_j -
2 \partial_a n_i \partial_b n_i \partial_a n_j \partial_b n_j )$,
associated with spin wave scattering; this is described by the RG flow
$d \lambda/ d \ell = (d-2)\lambda + c \rho_s / M_0 $
(with $c$ a positive constant), which sets in at scales $>\xi_0$.
For $d<2$, $\lambda$ flows to a fixed point value
$ c\rho_s/(2-d)M_0$. Similar phenomena are expected for other, even less
relevant, interactions. Thus all the irrelevant couplings actually flow
either to zero or to nonzero fixed point values, and approach these values
with eigenvalues given by their dimensions established above, $y_k=d+2-k$.
The simple form of these results, compared with more familiar field theories,
is due to the fluctuationless fully-polarized ground state, so that
contributions come only from scattering of already-existing spin waves
(similar to the dilute Bose gas \cite{sss}), and to the rotational symmetry
requirements.

For $d<2$, these considerations imply that
all observables should be universal functions of the bare couplings $M_0$,
$\rho_s$ and $H$, realizing
a {\em no-scale-factor universality\/} similar to that discussed in
Ref~\cite{sss} for
the dilute Bose gas in $d < 2$. Scaling forms can therefore be deduced from
a naive dimensional analysis of the length and time scales in the CQFM; for
the free energy density $F$ we obtain
\begin{equation}
F =  T M_0 \Phi_F \left({\bar{\rho}_s}/{T},\, {H}/{T} \right)
\label{free}
\end{equation}
where $\bar{\rho}_s \equiv \rho_s / M_0^{(d-2)/d}$ is a rescaled stiffness
and $\Phi_F ( r, h)$ is a universal scaling function with
no arbitrary scale-factors and dependent only on $d$ and the symmetry group
($O(3)$) of the
ferromagnet (in particular, for the lattice ferromagnet, $S$
enters only indirectly, through $\rho_s$ and $M_0$).
Scaling forms for other thermodynamic observables can be obtained by taking
derivatives of $F$. Because the scaling form is obtained by setting the
irrelevant couplings to their fixed point values, it is valid only when the
deviation of those couplings from their fixed points are negligible.
If the bare values of the irrelevant couplings (defined at the scale $\xi_0$)
are set to their fixed point values in (\ref{cqfm}), then the free energy is
given by (\ref{free}) at all values of $T$, $\rho_s$, $H$. For a real system,
this tuning of parameters does not occur, and the behavior approaches the
universal form only for $T<T_{\mbox{max}}(H)$ or $H<H_{\mbox{max}}(T)$;
we expect $T_{\mbox{max}}\sim\bar{\rho_s}$ as $H\rightarrow 0$ for systems with
small $S$.
For $d>2$ additional scaling variables, associated with other relevant
couplings, will be necessary in a generalization of (\ref{free}).

We now consider the different $T$ regimes of the CQFM, ignoring ${\cal L}_1$,
so this will be universal for $d<2$ and marginally so for $d=2$.
Fig~\ref{phasediag} shows a phase diagram as a function of the three
dimensionless
ratios of the energy scales $T$, $H$, and $\bar{\rho}_s$ plotted in
the projective plane.
All boundaries are smooth crossovers with the exception (for $d > 2$) of the
ferromagnetic phase transition at the single point $H=0$, $T=T_c \sim
\bar{\rho}_s$.
The regimes in Fig~\ref{phasediag} are: \\
({\em i\/}) {\em Quantum activated} (QA), $T < H$---most spins are aligned
as in the ground
state along $H$, with thermal corrections associated with a thermal
activation factor
$e^{-H/T}$.
There is also a crossover (indicated by the
dotted line) between $T < \bar{\rho}_s < H$ and $\bar{\rho}_s < T < H$,
but it only affects the pre\-factor of $e^{-H/T}$. \\
({\em ii\/}) {\em Renormalized classical} (RC), $H < T <
\bar{\rho}_s$---the behavior is dominated by fluctuations of classical
Goldstone modes
with energies smaller than $T$. \\
({\em iii\/}) {\em Quantum critical} (QC), $T > H$, $\bar{\rho}_s$---this
regime
was proposed recently in $d=2$ in Ref~\cite{sokol}. It is the high $T$ limit
of the CQFM\@. One may interpret
the behavior here as the response of the $M_0\neq0$ system with zero
Hamiltonian to a finite size, $1/T$, along the time direction.
Ref~\cite{sokol} also suggested that, for $S=1/2$, $T_{\mbox{max}}$ is
large enough for the square lattice ferromagnet to exhibit QC behavior.

In $d=1$, Nakamura and Takahashi~\cite{taka} have studied the magnetization
of the
spin $S$ chain in the RC region, and their results are described by the CQFM\@.
The expected scaling form is $M= M_0 \Phi_M$, where $\Phi_M$ is a function
similar
to $\Phi_F$; they find a scaling function,
$\phi_M$, to which our function $\Phi_M$ reduces in a limit appropriate for
the RC region: $\Phi_M ( r \rightarrow \infty , h \rightarrow 0 ) =
\phi_M (rh)$ and they computed $\phi_M (y) = 2y/3 - 44 y^3 / 135 +
\ldots$ for small $y$. For $d<2$ the function $\Phi_M (r,h)$ can also be
computed in the usual  spin wave expansion~\cite{dyson,kosevich} which yields a
universal series containing integral powers of $r^{-d/2}$ times functions of
$h$.

In $d=2$, the flow of $\lambda$ is logarithmic, and
universality at low $T$ is violated by logarithms, unlike the situation in
antiferromagnets \cite{CHN}.
As a result, the QC
regime lies at the edge of where quasi-universality holds.
However the logarithmic terms
contain pre\-factors of powers of $T$ at low $T$, and are absent in the
leading low $T$ behavior. The flow of $\lambda$, in particular, has been
overlooked in previous analyses of $d=2$ ferromagnets\cite{kc}.

Before turning to calculations of the scaling functions, we discuss
other aspects of the $d=2$ case more fully. For the CQFM in general in
$d=2$, a conserved topological
current, defined as $j_\mu=\epsilon_{\mu\nu\lambda}
\epsilon_{ijk}n_i\partial_\nu n_j\partial_\lambda n_k/8\pi$
($\mu=x$, $y$, $\tau$), exists and represents the
number density and current of skyrmions.
The skyrmions experience an effective orbital magnetic field of
strength $4\pi M_0$, produced by the Berry phase
term, as can be seen from the identity $M_0\int\vec{A}(\vec{n})
\cdot \partial_{\tau}\vec{n}=\int {\bf j}\cdot{\cal
A}$, where $\nabla\times{\cal A}=4\pi M_0$ represents the uniform field. Thus
there is an effective magnetic length for the skyrmions, that is related to
$\xi_0$; moreover skyrmions come in quantized sizes that are multiples of
$\xi_0^2$. In the quantum Hall system, skyrmions carry real electric charge
\cite{kane,sondhi}, so the use of the CQFM does not exclude charged
excitations. However, in this case, there are also other terms that are
known to appear in the long-wavelength description \cite{kane,sondhi} but
are not included in the CQFM thus far. The extra terms are
({\it i\/}) the Hopf term $2\pi i\int j_\mu a_\mu$, where $a_\mu$ obeys
$\epsilon_{\mu\nu\lambda}\partial_\nu a_\lambda =j_\mu$, which endows the
skyrmions with
Fermi statistics and half-integral spin \cite{nayak}; ({\it ii\/}) the
Coulomb interaction
$\int\int j_{\tau} (x)j_{\tau} (x')e^2/(2\varepsilon |x-x'|)$.
The Hopf term is marginal, but since it contains a time derivative, it affects
the quantization directly, and may change the dimension of the Hilbert space in
a finite system. It will not affect the discussion of universality and its
violation by logarithms, but it will change the precise scaling functions in
general. The Coulomb interaction has dimension 1, so is relevant, though less
so than $\rho_s$, and in principle requires that an additional scaling variable
appears in the scaling functions.
However, both terms enter only through skyrmions which, in
the large $\rho_s$ region, always have an energy $>\rho_s$, and their
contributions are exponentially small at low $T$.

Finally, we present our large $N$ results for the CQFM\@.
These are valid over the entire phase diagram of Fig~\ref{phasediag},
and exhibit all the crossovers. We discuss two
different large $N$ limits; the first generalizes the symmetry group from
$SU(2)\cong O(3)$ to $SU(N)$ and the second to $O(N)$. To obtain the
$SU(N)$ theory
we write $\vec{n} =z_{\alpha}^{\ast} \vec{\sigma}_{\alpha\beta} z_{\beta}$,
where
$\vec{\sigma}$ are the Pauli matrices, $z_{\alpha}$ is a 2-component
complex field,
and $\sum_{\alpha}|z_{\alpha}|^2 = 1$. The Berry phase in ${\cal L}$ now
becomes
$2 M_0 \sum_{\alpha}z_{\alpha}^{\ast} \partial_{\tau} z_{\alpha}$. We can
now obtain
$SU(N)$ symmetry  by allowing $\alpha$ to run from 1 to $N$ (for $N$ even,
the field $H$ is taken
to couple to the generator $\mbox{diag} (1_{N/2} , -1_{N/2})$); the
gradient terms
are as in the $CP^{N-1}$ model~\cite{kane,cpn}.
For the $O(N)$ generalization we parametrize $n_i = i\epsilon_{ijk}
w_{j}^{\ast} w_k$ where $w_i$ is a 3-component complex field obeying
$\sum_i |w_i|^2 =1$
and $\sum_i
w_i^2 = 0$. The Berry phase is now $M_0 \sum_i w_i^{\ast} \partial_{\tau}
w_i$ and $O(N)$
symmetry is achieved by allowing
$i$ to run from $1$ to $N$ (for $N$ divisible by $3$, $H$ couples
to a generator which
contains $N/3$ copies of the $O(3)$ generator).
The $1/N$ expansion of both theories is standard
and we omit all details: the constraints are imposed by Lagrange
multipliers, and $\rho_s$
and $M_0$ should be of order $N$ as $N\rightarrow\infty$. We present below
$N=\infty$ results from both theories for some observables in $d=2$
(although results can be
obtained for arbitrary $d$); the results are
universal as the logarithmic
violations of universality appear only at higher orders in $1/N$.\\
({\em a}) {\em Magnetization\/}: From the $SU(\infty)$ theory we obtain:
\begin{equation}
\Phi_M (r,h) = \ln \left[ (q_1-e^{-h/2})/(q_1-e^{h/2})\right]/(8 \pi r)
\label{sun}
\end{equation}
where $q_1>1$ is the solution of
$(q_1 - e^{-h/2} ) ( q_1 - e^{h/2} ) = q_1^2 e^{-8 \pi r}.$
Similarly we obtain from the $O(\infty)$ theory:
\begin{equation}
\Phi_M (r,h) = \ln \left[ (q_2-e^{-h})/(q_2-e^{h})\right]/(4 \pi r)
\label{on}
\end{equation}
where $q_2>1$ is the solution of
$(q_2 - e^{-h} ) (q_2 - 1) ( q_2 - e^{h} ) = q_2^3 e^{-4 \pi r}.$
We show in Fig~\ref{plotm} a plot of these results for
$M/M_0$ as a function of $T/H$ for a few values of $\rho_s /H$, including
$\rho_s/H=0$.
For $\rho_s
\gg H$ it is possible, in principle,
to use simpler functions characteristic of the different regions of
Fig~\ref{phasediag},
punctuated by crossovers between them. At the
lowest $T$ we have QA behavior with $\Phi_M - 1\propto e^{-h}$. At larger
$T$ we have RC
behavior described by the scaling function of Ref~\cite{bz};
in the $SU(\infty)$ theory we can obtain this function
from (\ref{sun}): $\Phi_M = 1 + \ln[h/2 + ((h/2)^2 + e^{-8\pi r})^{1/2}]/(4
\pi r)$.
At the largest $T$ we have $QC$ behavior in which we expect
$\Phi_M  \propto h$.
Although the analytic forms are rather different in the 3 regimes, the
qualitative
trends in an $M$ vs.\ $T$ plot are similar; this makes picking out the regimes
from experimental data rather difficult.\\
({\em b}) {\em NMR relaxation rate $1/T_1$\/}: Unlike the static
magnetization, the
dynamic susceptibility has significantly different behavior in the regions of
Fig~\ref{phasediag}, and this leads to clear signatures of them in $1/T_1$.
We model the
nuclear-electron
contact coupling by
$A M_0
\vec{I}(\tau) \cdot \vec{n} (0, \tau)$. Then the relaxation rate is given
by $1/T_1 = A^2 T
\lim_{\omega \rightarrow 0}
\mbox{Im} \chi_{L+-} (\omega) /\omega$ where $\chi_{L+-}$ is the local
transverse
susceptibility. Dimensional analysis shows that $1/T_1$
satisfies the scaling form $1/T_1 = (A^2 M_0^2 /T ) \Phi_{T_1}$,
where $\Phi_{T_1}$ is a universal function like $\Phi_F$.
We evaluated
$\Phi_{T_1}$ in both large $N$ limits and found
\begin{equation}
\Phi_{T_1} (r,h) = 1/[16 \pi r^2 (q_1 e^{h/2} -1 )]
\end{equation}
in the $SU(\infty)$ theory (with $q_1$ defined below (\ref{sun})) and
\begin{equation}
\Phi_{T_1} (r,h) = (1/8 \pi r^2 )[1/(q_2 e^{h} -1 ) + 1/(q_2 -1)]
\end{equation}
in the $O(\infty)$ theory (with $q_2$ defined below (\ref{on})).
A plot of
these results
for $1/T_1$ is shown in Fig~\ref{plott1} as a function of $T/\rho_s$ for some
values of
$\rho_s / H$.
The most notable feature is the  ``ferromagnetic coherence peak''
which signals a crossover between the QA and RC regimes.
This becomes clear from the asymptotic behavior for $\rho_s \gg H$.
In the low $T$, QA regime we have activated behavior $1/T_1 \sim e^{-H/T}$ (
$\Phi_{T_1} = e^{-h} /16 \pi r^2$
for $SU(\infty)$ and $\Phi_{T_1} = e^{-h} /8 \pi r^2$ for
$O(\infty)$). In contrast, in the RC regime, $1/T_1$ decreases
exponentially fast with
increasing $T$ (
$\Phi_{T_1} = e^{4\pi r} /16 \pi r^2$
 for $SU(\infty)$ and $\Phi_{T_1} = e^{4\pi r/3} /4 \pi r^2$ for
$O(\infty)$) due to the rapid decrease in the ferromagnetic correlation length;
this behavior of $1/T_1$ is similar to that observed in the RC region of
$d=2$ quantum antiferromagnets~\cite{imai}. Finally in the large $T$ QC region
we find $1/T_1 \sim \mbox{const}$ ($\Phi_{T_1} = 1/4r$ for $SU(\infty)$ and
$\Phi_{T_1} = 1/3r$ for $O(\infty)$). Notice from Fig~\ref{plott1}
that the coherence peak survives even for moderate values of $\rho_s / H$,
though it may be absent for $\rho_s/H$ sufficiently small.
We emphasize that for $\rho_s/H$ large, this peak is not dependent upon a large
value of $T_{\mbox{max}}$, as it occurs at the crossover between the low $T$
QA and RC regimes.

Nontrivial textures (skyrmions) exist, and the Hopf and Coulomb interaction
terms can be included, for all $N$ in both the $SU(N)$ and $O(N)$ models,
but have no effect in the $N\rightarrow\infty$ limit.
While agreement with the quantum Hall
experiments~\cite{sean} is fair, at this point it is not clear whether the
differences between theory and experiment are due to these effects, other
differences between $N=\infty$ and $N=2$ or $3$, the possibility that
$T_{\mbox{max}}$ is small, or the uncertainty in the value of $\rho_s$. More
complete measurements of the $T$ dependence of $1/T_1$, particularly in
samples with a larger $\rho_s / H$, could help answer these questions.

We thank S.E. Barrett and A.V. Chubukov
for numerous invaluable discussions, and acknowledge
helpful remarks by A. Sokol and S.L. Sondhi.
The research was supported by NSF Grants, Nos.\ DMR-92-24290 and DMR-91-57484.

\begin{figure}
\caption{Phase diagram of the CQFM as a function of
dimensionless
ratios of the 3 energies $T$, $H$ and $\bar{\rho}_s \equiv \rho_s /
M_0^{(d-2)/d}$.
Each energy becomes infinite at one of the vertices, and equals zero on the
opposite side.
Dashed and dotted lines are crossovers. Increasing $T$ from zero to
$\infty$ at fixed
$\bar{\rho}_s/H$ corresponds to moving along a straight line from the base
to the apex.  }
\label{phasediag}
\end{figure}

\begin{figure}
\caption{Magnetization of the CQFM in $d=2$, computed in the $N=\infty$ limit
of $SU(N)$ and $O(N)$ theories for a number of values of $\rho_s / H$,
and compared with
the experiments of
Ref~\protect\cite{sean}.
The $\rho_s=0$ limit of the large $N$ results yield
the spin ${\cal S}$ Brillouin function, with ${\cal S}=1/2$ in the $SU(N)$
model, and
${\cal S}=1$ for $O(N)$.}
\label{plotm}
\end{figure}

\begin{figure}
\caption{As in Fig.~\protect\ref{plotm} but for $1/T_1$ vs. $T/\rho_s$. The
constant
$A^{\prime}
= \rho_s /A^2 M_0^2 $.}
\label{plott1}
\end{figure}

\end{document}